\documentstyle[preprint,revtex]{aps}
\newcommand{\BE}{\begin{equation}}
\newcommand{\EE}{\end{equation}}
\newcommand{\BEA}{\begin{eqnarray}}
\newcommand{\EEA}{\end{eqnarray}}
\begin{document}
\draft

\begin{title}
Dynamic scaling and quasi-ordered states
in the two dimensional Swift-Hohenberg equation
\end{title}

\author{K.R. Elder$^{1}$, Jorge Vi\~nals$^2$ and Martin Grant$^{1}$}

\begin{instit}
$^{1}$ Department of Physics, McGill University,
Rutherford Building, 3600 University Street,
Montr\'eal, Qu\'ebec, Canada H3A 2T8,
$^{2}$ Supercomputer Computations Research Institute, B-186,
Florida State University, Tallahassee, Florida 32306-4052.
\end{instit}

\receipt{May 19, 1992}

\begin{abstract}

The process of pattern formation in the two dimensional Swift-Hohenberg
equation is examined through numerical and analytic methods.
Dynamic scaling relationships are developed for the collective ordering
of convective rolls in the limit of infinite aspect ratio.
The stationary solutions are shown to be strongly influenced
by the strength of noise. Stationary states for small and large noise
strengths appear to be quasi-ordered and disordered respectively.
The dynamics of ordering from an initially inhomogeneous state is very
slow in the former case and fast in the latter.
Both numerical and analytic calculations indicate that the slow
dynamics can be characterized by a simple scaling relationship, with a
characteristic dynamic exponent of $1/4$ in the intermediate time regime.

\end{abstract}


\narrowtext

\section{Introduction}

Convective instabilities in liquid systems provide an interesting
example of nonlinear, nonequilibrium processes.  An example of such phenomena
that has received considerable attention in recent years is the onset
and formation of roll patterns in Rayleigh-B\'enard convection
\cite{re:sw77,re:el92,re:el90,re:st84,re:me87,re:xi91,re:po79,re:po81,re:kr85,re:ah81,re:ne83,re:to81,re:sc86,re:vi91}.
In this paper we consider a simple
model of this process, the Swift-Hohenberg \cite{re:sw77} (SH) equation,
to study the situation in which a nonconvective state is
brought above the convective threshold, for a system of infinite
horizontal dimension.  Once above the convective threshold,
a complex pattern of convective rolls emerges.  By studying extremely
large systems, we focus our attention on the collective behavior of a
large number of rolls. If all sources of
fluctuations are small (let
$F$ denote their amplitude) the rolls form
small locally ordered domains (defined by regions of rolls with the same
orientation) that later reorient to become parallel to rolls of
larger domains in their immediate vicinity. Dynamic
scaling relationships analogous to those found in spinodal decomposition
are used to understand the transient dynamics of the collective ordering,
while ideas based on critical phenomena are used to analyze the asymptotic
steady states.

The SH equation was developed by considering the set of equations for a
simple fluid in the Boussinesq approximation, bounded by two infinite
horizontal plates separated by
a distance $d$, at temperatures $T_1$ and $T_1+\Delta T$ respectively.  For
values of the Rayleigh number $R$ larger than a critical Rayleigh number
$R_{c}$, an instability occurs giving
rise to convective rolls. The solution of the linearized equations
for the velocity field perpendicular to the plates $v_{z}$, and for the
temperature field $\theta$ (or more precisely for the deviations from
a linear temperature gradient between the plates)
contains stable and unstable eigenvalues.
The SH equation is obtained by neglecting all terms proportional to the
stable mode and by considering only the wavelengths near the most
unstable wavelength obtained in the linear analysis.  The SH equation is
asymptotically correct in the limit $R \rightarrow R_c$.
In dimensionless units, the SH
equation reads,
\BE
\label{eq:sw}
\frac{\partial \psi(\vec{r},t)} {\partial t} = \left( \epsilon -
\left(1+\nabla^2\right)^2\right)
\psi(\vec{r},t) - \psi^3(\vec{r},t) + \eta(\vec{r},t), \EE
where $\psi$ is a scalar, two-dimensionald field related to the
amplitude of the
eigenfunction corresponding to the unstable mode and is commensurate
with the convective rolls. The quantity $\epsilon = (R-R_{c})/R_{c}$ acts
as a control parameter, and $\eta$ is a random field, Gaussianly
distributed, with zero mean and correlations,
\BE
\label{eq:noise}
< \eta(\vec{r},t) \eta(\vec{r}',t')> = 2F
\delta(\vec{r}-\vec{r}') \delta(t-t'),
\EE
where $F$ is the intensity of the noise.

The salient feature of the SH equation is that the Lyapunov functional
associated with Eq. (1) is minimized by a one dimensional periodic
function which corresponds to a configuration comprised of
straight, parallel rolls. The appearance of such \lq\lq striped"
patterns is a common feature in
nature and has also been observed in magnetic films, block co-polymers,
the visual cortex, liquid crystals, microemulsions and eutectic growth.
Equation (\ref{eq:sw}) can be expressed in terms of a
Lyapunov functional ${\cal F}$ in the following manner,
\BE
\label{eq:eqfr}
\frac{\partial \psi(\vec{r},t)} {\partial t} = -
\frac{\delta {\cal F}(\psi)}{\delta \psi( \vec{r},t)} + \eta (\vec{r},t),
\EE
where ${\cal F}$ is given by,
\BE
\label{eq:lyap}
 {\cal F} = \int d\vec{r} \left[ \psi \left( \epsilon -
\left(1+\nabla^2 \right)^2 \right) \psi/2 + \psi^4/4 \right].
\EE
As noted in the original work of Swift and Hohenberg \cite{re:sw77}, this
model does not fall within the classification scheme of Halperin and
Hohenberg \cite{re:hh77}. Although other systems select a characteristic
wavelength which is finite (as in order-disorder or anti-ferromagnetic
transitions), this
model differs in that the pattern of rolls is rotationally invariant.
Thus small changes in orientation are associated with very small changes
of ${\cal F}$.
Another important feature of Eq. (\ref{eq:eqfr}) is
that $\psi$ is a non-conserved variable since $\int d\vec{r} \psi(\vec{r},t)$
can vary in time. In Section (II) a Lyapunov functional
with these two generic properties is considered,
in order to both study the SH equation and to establish the
extent to which our work may apply to other systems.

For the one-dimensional SH equation, Pomeau and Manneville \cite{re:po79}
have obtained
an approximate solution $\psi^{1d}$ that minimizes
${\cal F}$ in the limit of small $\epsilon$.
This solution will be used to estimate
the size of various quantities that enter into the analytic calculation
given in Section (II). The solution is given in a power series in
$\epsilon$, and is,
\BE
\label{eq:swst1d}
\psi^{1d}(x) = \sum_{i=0}^{\infty} a_i \sin \left( k_{0} \left[ 2i+1 \right]x
\right),
\EE
where, $ a_0(k_{0}) = \sqrt{ 4 \omega (k_{0})/3 }$,  $a_1(k_{0}) = -
a_0(k_{0})^3/(4 \omega (3k_{0}))$, and
$\omega (k_{0}) = \epsilon - (1-k_{0}^2)^2$, and $k_{0} = 1 - \epsilon^2/1024
$.
In two dimensions, a solution that minimizes
${\cal F}$ is for example $\psi(x,y)=\psi^{1d}(x)$. This solution corresponds
to a
set of straight rolls parallel to the $y$ direction. By analogy with a liquid
crystal, we
will refer to such a configuration as smectic since it
contains both long-range translational
and orientational order.
Although this solution minimizes ${\cal F}$, it is
doubtful that it could be obtained via numerical simulation of the SH
equation from an initially random state due to the difficulty in removing
defects in the absence of fluctuations. In
fact, Pomeau and Zaleski \cite{re:po81}, and Kramer and
Zimmermann \cite{re:kr85} have derived
other stationary solutions to the one-dimensional
SH equation that are not a global
minimum of ${\cal F}$ and involve a local shift in the phase
of $\psi$ (e.g., $\psi ~ sin(k_ox)$ for $x<-a$ and
$\psi ~ sin(k_o x + \pi)$ for $x> a$).

The presence of fluctuations (i.e., $F \neq 0$) changes matters
significantly. In one dimension long
range order is broken. A numerical solution of the stochastic
SH equation in one dimension by Vi\~nals {\it et al.} \cite{re:vi91}
has shown that a finite $\epsilon'$ leads to a diffuse peak in
$S(k,t)$. This one
dimensional result has important implications for the two dimensional
stationary state. If the two dimensional stationary state is a set of
parallel rolls, then no long range translational order can exist, since the
one dimensional solution should be valid normal to the rolls.
The two dimensional equilibrium state could still, however, exhibit long range
orientational order.  One interesting feature of the SH equation
is that the
orientational order parameter (i.e., a vector that lies normal
to the convective rolls) is continuous, similar to the spin field in
the XY model. Of course the distance between spins in the XY model
is fixed on a lattice, whereas in the SH equation the distance between
convective rolls can vary. Since the XY model is the
prototypical model of a Kosterlitz-Thouless \cite{re:ko73}
transition it is possible that
the SH model contains such a Kosterlitz-Thouless phase. In fact Toner
and Nelson \cite{re:ne83,re:to81} have suggested that this could be
the case. In
Sections (II) and (III) a more detailed discussion of the
stationary states will be presented.

The dynamic evolution to the stationary states is
realized through the enlargement of regions of convective rolls with
the same orientation and the elimination of defects.  A singular
perturbation solution \cite{re:el90} to the SH equation predicts that
correlations in
$\psi$ obey the following scaling relationship,
\BE
\label{eq:scsw}
S(k,t) = t^{x} f([k-k_{0}]t^x)
\EE
where, $S(k,t) = \sum_{\hat{k}} <|\psi(\vec{k},t)|^2>/\sum_{\hat{k}}$,
$k$ is the wavevector, $x$ is a dynamic scaling exponent,
and the orientation of $\hat{k}$ have been averaged over.
This relationship should be valid
for extremely large systems, comprised of many domains
of different orientation.
Equation (\ref{eq:scsw}) is analogous to
the scaling relationships observed in spinodal decomposition and
order-disorder transitions.  In domain growth phenomena the structure factor
has been found to obey the following dynamic scaling
relationship \cite{re:gu83,re:oo88},
$S(k,t) = R(t)^d f \left[(kR(t) \right]$, where $R(t) \propto t^x$
is the average domain size and $d$ is the dimension of the system.
The differences between this result and Eq. (\ref{eq:scsw}) are
due to the existence of a non-scaling length: the roll width.
The specific value of $x$ has been shown to be of considerable
importance in first order phase transitions, and is particular to a
given universality class. Systems with a conserved or
non-conserved scalar order parameter fall into classes
characterized
by $x=1/3$ and $x=1/2$ respectively.

In Sections (II) and (III) analytic and numerical methods are used to
estimate the dynamic exponent $x$ for the
SH equation. The analytic calculation is an expansion around a curved set of
rolls of varying width, and is an extension of the interfacial methods used
in first order phase transitions. Some of the results of Toner and
Nelson \cite{re:ne83,re:to81} alluded to in the
preceding paragraph are recovered in Section (II), and are shown to
be consistent with the numerical results given in Section (III).
In the final section a discussion of these results is presented and
comparisons with other systems is made.

\section{Dynamics of Curved Rolls}

Insight into the ordering dynamics can be obtained by considering
the relaxation of a set of convective rolls in which the
orientation and spacing varies slowly in space.
This description precludes the explicit consideration
of defects, which will be discussed later.
A typical pattern that falls within this description is shown in
Fig. (\ref{fi:1}) which was taken directly from the numerical
calculations to follow. In the absence of defects, the dynamic evolution
can roughly be separated into three distinct mechanisms:
the relaxation of the local curvature of the rolls, the relaxation of the
distance between consecutive
rolls, and the relaxation in the functional form of $\psi$ (to be described
later).
The latter fluctuations will be seen to  decay much faster than the
other mechanisms which describe changes in the positions of the
roles. To begin the calculation a Lyapunov functional ${\cal F} \{ \psi \}$ of
a scalar function $\psi$ of the form,
\BE
\label{eq:gf}
 {\cal F} \{ \psi \}
= \int d\vec{r} \left[ -\psi \left(f\left(\nabla^2\right)\right)
\psi/2 + g(\psi) \right],
\EE
is considered, with the restrictions that ${\cal F}$ is minimized (in
the absence of noise) by a roll structure with characteristic
wavenumber $k_{0}$,
$\psi$ is non-conserved and $g$ is an even function of $\psi$.
Subject to these
restrictions the functions $f$ and $g$ can be written,
$f = \sum_{i=0} a_i (\nabla^2)^i$
and $g = \sum_{i=1} c_i (\psi^2)^{i+1} $.
The SH equation can be recovered by choosing
$a_0=\epsilon -1, a_1 = -2, a_2=-1$ and $a_{i>3} =0$,
and $c_1 = 1/4$ and $c_{i>1} =0$. We restrict our analysis to
even powers of $\psi$
due to the symmetry of roll patterns, while pointing out that odd powers of
$\psi$ can lead to interesting behavior such as stationary solutions
with hexagonal symmetry.
The dynamical evolution of $\psi$ is
given by,
\BE
\label{eq:gem}
\frac{\partial \psi}{\partial t} = - \frac{\delta {\cal F}}{\delta \psi}
+ \eta (\vec{r},t) = f(\nabla^2)\psi - \delta g/\delta \psi + \eta
(\vec{r},t).
\EE
Other systems that are modeled by Eqs. (\ref{eq:gf}) and (\ref{eq:gem})
include pattern formation in the visual
cortex \cite{re:th91} and a model of uniaxial ferromagnetic
films \cite{re:ro90}.

In order to describe
the relaxation of the configuration illustrated in Fig. (\ref{fi:1}),
the Cartesian coordinates $x$ and $y$ are mapped onto
curvlinear coordinates $u$ and $s$ as indicated in the figure.
In this new coordinate system, a vector
$\vec{r}(x,y)=x\hat{x}+y\hat{y}$
becomes $\vec{r}_m(s,u)=\vec{R}_m(s)+(u_m-2m\pi/k_{0})\hat{n}_m(s)$,
where $\vec{R}_m(s)$ is the
location of the $m^{th}$ roll, $\hat{n}_m(s)$ is the normal
to the $m^{th}$ roll, $u_m$ is restricted to
the region $(2m-1)\pi/k_{0} < u < (2m+1)\pi/k_{0}$,
and $m=0, \pm 1, \pm 2 , ... $.
The coordinate system is shown in Fig. (\ref{fi:1}).
In the new coordinates the Laplacian operator becomes,
\BE
\label{eq:lap}
\nabla^2 = \frac{\partial^2}{\partial u^2}
+\frac{\kappa}{(1+(u_m-m\pi/k_{0})\kappa)}\frac{\partial}{\partial u}
+\frac{1}{(1+(u_m-m\pi/k_{0})\kappa)^2}\frac{\partial^2}{\partial s^2}
-\frac{(u_m-m\pi/k_{0})\kappa_s}
 {(1+(u_m-m\pi/k_{0})\kappa)^3}\frac{\partial}{\partial s}
\EE
where the curvature $\kappa = - \partial\theta(s)/\partial s$, $\kappa_s =
\partial \kappa/\partial s$ and $\theta(s)$ is the
angle between the tangent to the $u_m=2m\pi/k_{0}$ curve and the $x$ axis.
Assuming
that the curvature of all the individual rolls
is small, i.e., $\pi\kappa/k_{0} << 1$, Eq. (\ref{eq:lap})
becomes,
\BE
\label{eq:lapa}
\nabla^2 \approx \frac{\partial^2}{\partial u^2}
+\kappa \frac{\partial}{\partial u}
+\frac{\partial^2}{\partial s^2}.
\EE
Coupling between rolls is incorporated by considering the fluctuations
in the separation between them. The field $\psi$ is expanded
around the one dimensional stationary solution in the following manner,
\BE \label{eq:psia}
\psi(\vec{r},t)=\psi^{1d}(u_m(\vec{r},t)+h(u_m,t)) + \delta \psi(\vec{r},t),
\EE
where, $\psi^{1d}$ is defined by the equation,
\BE
\label{eq:psis}
f(\partial^2/\partial u^2) \psi^{1d}
-\frac{\delta g(\psi)}{\delta \psi}|_{\psi^{1d}} = 0.
\EE
This decomposition describes the relaxation of the local curvature
through the
curvature dependence of the Laplacian (Eq. (\ref{eq:lapa})),
and the relaxation
of the separation between rolls through the function $h$. $\delta\psi$ takes
into account fluctuations about the one dimensional functional form of $\psi$
described by Eq. (\ref{eq:psis}).
The validity of this expansion depends
on the assumption that the one-dimensional solution $\psi^{id}$
is valid normal to the lines defined by $\psi=0$.
The numerical results to follow verify that this is quite a good
approximation at late times. As can be seen in
Fig. (\ref{fi:1}),
$\psi^{1d}$ is found to be a good approximation to $\psi$ in the
direction normal to the rolls for distances of 6 to 20 consecutive wavelengths.
The assumption that $h$ is small precludes a rapid change in the phase
of $\psi$ that would be associated with defects.

We next substitute Eqs. (\ref{eq:lapa}) and (\ref{eq:psia}) into
Eq. (\ref{eq:gem}),  and expand to lowest order in
$\kappa$, $h$, $\delta \psi$, and their derivatives. This gives the result,
\BEA
\label{eq:eqw}
(\partial \psi^{1d}/\partial w)(\partial w/\partial t)
+ (\partial \delta\psi/\partial t)
 = &\ & \kappa \sum_{i=1}^{\infty} i a_i
 (\partial /\partial w)^{2i-1} \psi^{1d}
+ \kappa_{ss} \sum_{i=2}^{\infty} \frac{i(i-1)}{2!} a_i
(\partial /\partial w)^{2i-3} \psi^{1d} \nonumber \\
&+& \partial h/\partial u \sum_{i=1}^{\infty} (2i)a_i
(\partial/\partial w)^{2i} \psi^{1d} \nonumber \\
&+& \partial^2 h/\partial u^2 \sum_{i=1}^{\infty} \frac{(2i)(2i-1)}{2!} a_i
(\partial/\partial w)^{2i-1} \psi^{1d} \nonumber \\
&-& \left[\left(\delta^2 {\cal F}/\delta \psi^2\right)|_{\psi=\psi^{1d}}\right]
\delta \psi + \eta + \cdots
\EEA
where $w = u_m+h(u_m)$.
The coefficients of the terms proportional to $\kappa$ and
$\partial h/\partial u$ are identically zero if $\psi^{1d}$ is of
the form $\sin(k_{0}w)$ or $\cos(k_{0}w)$ with $k_{0}$ defined by
$(\partial f(k)/\partial k)|_{k=k_{0}} = 0$.  For the SH equation replacing
$\psi^{1d}$ by a sinusoidal function is an
extremely good approximation for small $\epsilon$ as was shown by
Pomeau and Manneville \cite{re:po79}. The coefficient of the term
proportional to $\delta \psi$ is less than zero since $\psi^{1d}$ is a
minimum of ${\cal F}$ which implies
$(\delta^2 {\cal F}/\delta \psi^2)_{\psi=\psi^{1d}} > 0$.

The evolution of $\delta \psi$ can be obtained by introducing
the projection operator, \\
${\cal P}_H \equiv k_o/(2\pi)\int_{(2m-1)\pi/k_{0}}^{(2m+1)\pi/k_{0}} dw$.
Applying ${\cal P}_H$ to Eq. (\ref{eq:eqw}) gives,
\BE
\label{eq:bulkm}
\partial \delta \psi/\partial t = \left(f(\nabla^2)
- {\cal P}_H (\delta^2 g(\psi)/\delta \psi^2)|_{\psi=\psi^{1d}} \right)
\delta \psi + \eta',
\EE
where, $<\eta'(n,s_1,t_1)\eta'(m,s_2,t_2)> = 2F
(\delta_{n,m}/(2\pi/k_{0}))
\delta(s_1-s_2) \delta(t_1-t_2)$.
In deriving this equation, it was explicitly assumed that the length scale
over which $\delta \psi$ varies is much greater than the roll wavelength,
that is,
$\delta \psi$ is approximately constant over length scales of the
order $2\pi/k_{0}$.
In addition it was assumed that the coefficient of $\partial h /\partial u$
was negligible.
For the SH equation, Eq. (\ref{eq:bulkm}) becomes,
\BE
\label{eq:bulksh}
\partial \delta\psi/\partial t =
\left(-2\epsilon - (1+\nabla^2)^2 \right)\delta \psi + \eta',
\EE
if Eq. (\ref{eq:swst1d}) to lowest order in $\epsilon$ is used for $\psi^{1d}$.
Thus $\delta \psi$ decays exponentially in time since
$-2\epsilon-(1-k^2)^2$ is less than zero for all $k$.
In the next few paragraphs, it will be shown that the other mechanisms
decay as a power law in time and consequently the
relaxation of the initial roll pattern considered is
controlled by the motion of the rolls.

These dynamics can be extracted by applying a different
projection operator, i.e.,  \\
${\cal P}_P \equiv k_o/(2\pi)\int_{(2m-1)\pi/k_{0}}^{(2m+1)\pi/k_{0}}
dw (\partial \psi^{1d} /\partial w)$.
Applying ${\cal P}_P$ to Eq. (\ref{eq:eqw}) gives,
\BE
\label{eq:intm}
\partial w/\partial t =
 b_1 \kappa  + b_2 \kappa_{ss}
 + b_3 \partial^2 h/\partial u^2  + \zeta(u,s,t)
\EE
where,
\BE
b_1 = \sum_{i=1}^{\infty} i a_i \frac{\sigma_{2i-1}}{\sigma_1},
\EE
\BE
b_2 = \sum_{i=2}^{\infty} a_i \frac{i(i-1)}{2!}
\frac{\sigma_{2i-3}}{\sigma_1},
\EE
\BE
b_3 = \sum_{i=1}^{\infty} a_i \frac{(2i)(2i-1)}{2!}
\frac{\sigma_{2i-1}}{\sigma_1},
\EE
\BE
\label{eq:sigs}
\sigma_i = k_o/(2\pi) \int_{(2m-1)\pi/k_{0}}^{(2m+1)\pi/k_{0}}
dw (\partial \psi^{1d}/\partial w)
(\partial^{i} \psi^{1d}/\partial w^i),
\EE
and $<\zeta(m,s,t) \zeta(n,s',t') > =
(2F/\sigma_1) (\delta_{m,n}/(2\pi/k_{0}))
\delta(s-s')\delta(t-t')$.
Equation (\ref{eq:intm}) is the main result of this calculation.
A simple dimensional analysis of Eq. (\ref{eq:intm}) reveals that two
different mechanisms provide different relaxational rates:
The distance
between rolls relaxes at a rate of $t^{-1/2}$, while the curvature of the
rolls relaxes at a rate of $b_1 t^{-1/2}+ b_2 t^{-1/4}$.
If $\psi^{1d} = A \sin (k_{0} u)$, then
the coefficient $b_1$ is identically zero, and the
rolls straighten at a rate of $t^{-1/4}$, which is considerably slower than
the relaxation of the fluctuations in the distance between rolls.
Therefore curvature relaxation is the dominant mechanism in the limit
$b_1=0$.
Corrections to a sinusoidal $\psi^{1d}$ are generally small close to
onset. This leads to
a small but finite value of $b_1$ (e.g., for $\epsilon=0.25$,
$b_1 \approx
1.0 \times 10^{-3}$). The low value of $b_1$ indicates that the asymptotic
behavior in which the curvature relaxes with a single power law
with an exponent
of $1/2$ should not be expected until very late times (i.e., of the
order $t=10^6$ for $\epsilon=0.25$).

Equation (\ref{eq:intm}) also describes the local fluctuations in the
position of the rolls.  If the magnitude of the noise
intensity (i.e., $2F/\sigma_1$) is
sufficiently strong, i.e., greater that the distance between
the rolls, the pattern of parallel rolls would be broken.
Hence a transition to a disordered state will occur when
$F_{KT} \propto \sigma_1/k_o^2$ (for the SH equation
this corresponds to $F_{KT} \propto \epsilon$).  It should be
noted that $\sigma$ is typically proportional to $k_o^2$.
The constant of proportionality is difficult to obtain, moreover it
would depend on the existence of defects.

	The effect of defects in related models has bee studied by
Toner and Nelson \cite{re:ne83,re:to81}.  Their analysis begins with
a generic free energy of the form,
\BE
\label{eq:tnf}
G(w) \equiv (B/2) \int d\vec{r} \left[
\left(\partial w/\partial y\right)^2
+ \lambda^2\left(\partial w^2/\partial x^2\right)^2 \right].
\EE
Assuming that $b_1$ is negligible,
Eq. (\ref{eq:intm}) can be written in the form,
\BE
\partial w(x,y,t) / \partial t  \approx -\delta G(w) /\delta w + \zeta.
\EE
where the free energy $G(w)$ is given by Eq.  (\ref{eq:tnf}), with $B=b_3$ and
$\lambda^2 = -b_2/b_3$. Toner and Nelson \cite{re:ne83,re:to81} have shown
that defects in conjunction
with the generic free energy given in Eq. (\ref{eq:tnf})
leads to algebraic decay in the orientational
correlation function ($C(r)$) , i.e.,
$C(r) = <e^{2i\phi(\vec{r})} e^{ 2i\phi(0)} > \approx |\vec{r}|^{-v}$, where
$\phi \equiv -\partial w / \partial x $ and \cite{re:waffle}
$ v = -2F/(\pi b_2 \sigma_1)$.  The algebraic decay with distance with a $F$
dependent exponent implies that the
state is a Kosterlitz-Thouless phase.  By using the
theory of Kosterlitz and Thouless \cite{re:ko73} a transition to a disordered
state occurs when $v = 1/4$ or $F_{KT} = -\pi b_2 \sigma_1 /8$.
For the SH equation (assuming $\psi^{1d} = \sqrt{4\epsilon/3} \sin (w)$) this
relationship becomes, $F_{KT} = \pi\epsilon/12$.
If an analogy with liquid crystals can be made, as suggested by Toner
and Nelson \cite{re:ne83,re:to81}, various phases can be identified;
an {\it isotropic} phase with short-range order
(for $F>F_{KT}$), a {\it nematic} or Kosterlitz-Thouless
phase with \lq\lq quasi" long-range orientational order (for
$0 < F < F_{KT}$) and a {\it smectic} phase with
long-range orientational and translational order (at $F=0$).
Portions of typical final configurations
along with the phase diagram are shown in Fig. (\ref{fi:2}).
An implication of this result is that in order to
observe a scaling relationship an orientational correlation
function should be measured,
however $S(k,t)$ should provide an accurate description of the
ordering dynamics when
the translational correlation length is larger than the average domain size.

The results of the preceding paragraphs indicate that the dynamic
in the absence of defects is controlled by the straightening of rolls,
and the rate associated with this mechanism is $t^{-1/4}$ during an
intermediate time regime.  The effect of defects on the
dynamics has been considered for a similar model (the XY model) by
Kawasaki \cite{re:ka85} and Loft and DeGrand \cite{re:lo87}.  Dimensional
arguments
and numerical simulations of the XY model led Kawasaki \cite{re:ka85} and Loft
and DeGrand \cite{re:lo87} respectively to the conclusion that the separate
mechanism of defect recombination occurs at
a rate of $t^{-1/2}$.  If the annealing away of defects in the SH equation is
the same as in the XY model then the curvature relaxation mechanism should
still
dominate in the intermediate time regime since it is the slower mechanism.
The expected crossover to $x=1/2$ at late times might however be influenced by
defects.
The more important influence of defects is the breaking of translational order
and creation of a disordered state above $F_{KT}$.

The idea of expanding $\psi$ around an almost parallel roll
configuration has been used by others, including Toner and
Nelson\cite{re:to81} and Ahlers {\it et al.} \cite{re:ah81}.  In these works
$\psi(x,y)$ is expanded around $\sin(k_{0}x)$.  Our work differs in that
$\psi$ is expanded around the one dimensional solution of the SH equation
(which is not in general a sinusoidal function) in a coordinate system
commensurate with a curved pattern of rolls.
In our work $\psi^{1d}$ is assumed to hold in the normal direction
to the line defined by $\psi=0$,
not simply in the $x$ direction.  In addition, the projection operator
techniques not only decouple fluctuations in the
functional form of $\psi$ from curvature and wavelength relaxation,
but also show that the fluctuations in the functional form
of $\psi$ are irrelevant since their decay is extremely fast.

\section{Numerical Solution}

The numerical results
to be presented were obtained by discretizing both space and time
derivatives in Eq. (\ref{eq:sw}).  Euler's method was used to discretize the
time derivative and the approximation for the Laplacian
included contributions from nearest and
next-nearest neighbors.
Schematically the numerical algorithm can be written,
\BEA
\label{eq:dsw}
\psi(i,j,n+1) = &\ & \psi(i,j,n)+ \Delta t \bigg[ \left( \epsilon -
[1+\nabla^2 ]^2 \right) \psi(i,j,n)  \nonumber \\
&-& \psi(i,j,n)^3
+ \eta(i,j,n) \bigg],
\EEA
where, the indices $(i,j)$ represent the coordinates $(x,y)$ and
the index $n$ represents time.  The Laplacian is evaluated using the
following discrete operator,
\BE \nabla^2 \psi(i,j) = \left( \frac{1}{4}\sum_{(nn)}+ \frac{1}{2}\sum_{(nnn)}
-3 \right)\psi(i,j)/(\Delta x)^2, \EE
where, the notation $(nn)$ and $(nnn)$ refers to the
nearest and next nearest neighbors to the site $(i,j)$ respectively.
Absolute time and
spatial coordinates are recovered by the simple relationships,
$t=n \Delta t$ and $\vec{r} = (i \hat{x} + j \hat{y})\Delta x$.

As in most numerical simulations of this nature the choice of
$(\Delta t,\Delta x)$ is dictated by the conflicting constraints
imposed by the need for numerical accuracy and the finite
computational power
available.  The former constraint requires $(\Delta t,\Delta x)$ to
be vanishingly small, while the latter requires the opposite.  In
practical terms the size of $\Delta x$ is limited by the smallest length
scale in the problem.  Typically the choice for $\Delta x$ imposes a
restriction on $\Delta t$ which can be obtained from a linear stability
analysis of the discrete map Eq. (\ref{eq:dsw}).  In some problems there may
also exist
non-linear numerical instabilities that further restrict the size of
$\Delta t$.  For the SH equation
$\Delta x$ must be smaller than the wavelength selected, which is of the
order $2\pi$. $\Delta x$ was chosen to be $2\pi/8$ to satisfy this
constraint. The linear solution of the discrete map given in
Eq. (\ref{eq:dsw}) contains a numerical instability (i.e., subharmonic
bifurcation) that occurs when,
\BE \Delta t > \frac{2 (\Delta x)^4}{(4-(\Delta x)^2)^2
-\epsilon(\Delta x)^4}.\EE
Thus to avoid numerical instability in the linear analysis
for $\epsilon=0.25$ and $\Delta x = 2\pi/8$, $\Delta t $ must be less
than 0.067 (for the standard nearest neighbor
discrete Laplacian operator  $\Delta t$ should be less than 0.014).
Although this analysis does not take into account the noise term or the
non-linear terms, it does
provide a minimum requirement.  Under these considerations a
time step of $\Delta t = 0.05$ was chosen and no numerical instabilities
were observed. Test runs with a smaller
$\Delta x$ indicate that the dynamics of $S(k,t)$ are insensitive to
decreasing grid size near $k=k_{0}$ \cite{re:fo2}.
The bulk of the simulations were run on a $512 \times 512$ periodic lattice
for $\epsilon=0.25$ at $F'=0,0.05$ and  $0.075$
(where $F'=F/(\Delta x)^2$) and averaged over
25 independent runs.  Smaller systems were used to examine the dynamics
(at $F' = 0.025, 0.065$ and $0.1$) and stationary solutions
(at a large number of values $F'$ between $0$ and $0.09$).
For non-zero $F'$, $\psi$ was initially set to zero, while for
$F'=0$, $\psi$ was chosen to be a random variable Gaussianly
distributed with zero mean and variance $0.1$.

The results of the numerical simulations suggest that there is
a qualitatively different behavior between the runs conducted at high and
low noise strengths.  At high noise strengths a steady state was rapidly
reached which corresponded to a disordered structure.  Figures (3a) and
(3b) display the configurations at $t=2500$ and $10^4$ respectively for
$F'=0.075$.  The striking similarity of these configurations and
small domain sizes indicates that a disordered stationary state has
been achieved.  In contrast, at low noise strengths a slow ordering of
domains was observed for all times probed.  Figures (3c), (3d) and
(3e) display the configurations at $t=10^2$, $10^3$ and $10^4$
respectively for $F'=0.05$.  A similar time sequence is shown
for $F'=0$ in Figs. (3f), (3g) and (3h).   It is important to
note that these runs had not reached a steady states as the rolls
were continuing to order.  The pronounced difference between
the low and high noise strength simulations was also apparent in
the structure factors and one-point distribution function
$\rho(\psi)$.  In Fig. (4a) $S(k,t=700)$ is shown for $F'=0.075$.
This structure factor was statistically indistinguishable from any
$S(k,t)$ past $t=200$.  The solid line in this figure is a fit to
$A/(B+(k^2-k_{0}^2)^2)$, which is suggestive of a disordered phase.
The structure factors shown in Fig. (4b) and (4c) for $F'=0.05$ and
$0$ respectively are significantly different for two reason.
First their shape is much sharper, and secondly they are continuing
to evolve in time.  Further indication of the qualitative differences
is shown in Fig. (5) which
compares $\rho(\psi)$ at the latest times for the three noise
strengths. The low noise strength one-point distributions are bimodal,
while the high noise strength $\rho$ is peaked at $\psi=0$.
To obtain an estimate of where the transition occurs at for $\epsilon=0.25$,
a stationary solution at $F'=0.09$ was obtained and
then $F'$ was decremented
in steps of $0.005$ at time intervals of $10^3$.  This test reveals a
transition in the region $F'_{KT} = 0.065-0.070$ which is
signaled by a sharp decrease in the peak height of $S(k)$, and a crossover
from a bimodal to unimodal distribution in $\rho$.

Although the numerical picture is consistent with the predictions
given in Section (II), it is important to note the limitations of the
numerical results.  The predictions of smectic, nematic and disordered
states were not unambiguously verified.  What is apparent is
that the runs above $F'_{KT}$ are disordered states.
The numerical results also strongly suggest that the
states below $F'_{KT}$ are qualitatively different from
those above.  Indeed,
the structure factors for the smectic and nematic simulations
were indistinguishable at the latest times.  If the $F'=0.05$ is
a nematic state, one should observe only quasi-long range orientation
order.  In contrast the numerical results indicated growth of translational
order up until the latest times probed.  Thus the numerical results
indicate a transition from a disordered state to one with long or
quasi-long range translational order.  Presumably, if later times were
probed an orientational order parameter would have to be introduced
to observe the subsequent ordering of domains.   At the very least the
transition observed corresponds to one in which there is a dramatic
change in translational order at $F'_{KT} \approx 0.0675$.

The dynamics of the collective ordering of rolls was analyzed
in terms of the dynamic scaling relationship given in Eq. (\ref{eq:scsw})
for $F' < F'_{KT}$.  In Figs. (\ref{fi:6}a) and (\ref{fi:6}b)
$S(k,t)$ is displayed for $F'=0$ and $0.05$ respectively at several
times.  The ordinate in these plots is $S(k,t)/t^x$ and the abscissa is
$(k_{0}-k)t^x$. If the scaling predictions are correct, the scaled form
of $S(k,t)$
should overlap for all times, near $k=k_{0}$.  For $F'=0$
(see Fig. (\ref{fi:6}a))
the structure factor scales remarkable well over the time range $t=10^2$ to
$10^4$, with a growth exponent of $x=1/5$.  Figure (\ref{fi:6}b) shows that the
dynamic scaling
hypothesis also works very well for $F'=0.05$, if a growth exponent of
$1/4$ is used.  If the scaling hypothesis is valid, all length scales
(except the convective roll width) should scale with the same dynamic
exponent. To investigate this hypothesis the height ($A$),
width ($w$) and various moments $m_n(t)$ of $S(k,t)$ were calculated.
The height and width were determined by fitting the top
portion of $S(k,t)$ to a
Gaussian of the form $Ae^{-((k^2-k_{0}^2)/w)^2)}$  and the moments
were defined to be $m_n(t) = \int_{k_{0}-w(t)}^{k_{0}+w(t)} dk |k-k_{0}|^n
S(k,t)$.  In fitting $S(k,t)$, $k_{0}$ was found to be very close to 1 which
is the value selected by linear theory.  The scaling relationship given in
Eq. (\ref{eq:scsw}) implies that $m_n(t) \propto t^{-nx}$,
$w \propto t^{-x}$ and $A \propto t^{x}$.
The results are shown in Figs. (\ref{fi:7}a) and
(\ref{fi:7}b) for $F'=0$ and $F'=0.05$ respectively.
For $F'=0$ the exponents measured from all lengths scales are
within two percent of $1/5$, while the exponents are all within one percent
of $1/4$ for
$F'=0.05$.  Although the statistics collected for $\epsilon=0.025$ and
$0.065$ were not sufficient to obtain accurate values of $x$, the measured
exponents were significantly closer to $1/4$ than $1/5$.

\section{Discussion and Summary}

The results of the numerical and analytic work provide several
predictions for the collective ordering of rolls in the limit of
infinite aspect ratio.  The stationary or steady states should
be strongly influenced by any random noise source, such that above
some critical noise strength (i.e., $F_{KT}$) an isotropic state
emerges, while below $F_{KT}$ an nematic or smectic state arises.
It should be emphasised that the numerical work indicates that the
transition is characterized by a drastic increase in translational order,
but was not sufficient to determine the precise nature of the states
below $F_{KT}$.  An independent estimate of the transition line
($F_{KT} \propto \epsilon$) was obtained in Section (II).
Further investigation at other values of $\epsilon$ would be
useful to establish this relationship.

The dynamical behavior of the collective ordering below
$F_{KT}$ was found to obey the dynamic scaling relationship
given in Eq. (\ref{eq:scsw}) for at least three decades in time.
In addition, a growth exponent of $x=1/4$ was observed in
the nematic region. This value of $x$ disagrees with an earlier
theoretical calculation and we have presented a possible resolution of
the discrepancy. We are, however, unable to explain the smaller exponent of
$x = 1/5$ observed for $\epsilon = 0$.
Although this work focused on the stochastic Swift-Hohenberg equation,
we expect that both features discussed, namely the existence of a
transition between a quasi-ordered to a disordered state, and the
asymptotic dynamical scaling behavior, are generic features of
two-dimensional systems defined by Eq. (\ref{eq:gem}).

The difficulty in observing these phenomena in Rayleigh-B\'enard
experiments is that previous experiments were conducted on
very small systems in which boundary effects play an important role.
In addition, estimates of the noise strength for real experiments is
typically much to small to observe the transition.  However recent
experimental studies of electrohydrodynamic convection in nematic liquid
crystals \cite{re:re91} have been able to detect and quantify the
amplitude of the fluctuations before onset (fluctuations in simple
liquids are believed to be too small to produce observable effects such
as the ones presented here). Furthermore, since typical roll widths in
these systems is of the order of microns, large aspect ratio samples are
readily available.  Experimental verification of the issues discussed in
this paper seems therefore possible.

\nonum
\section{Acknowledgments}
This international collaboration has been made possible by a grant from
NATO within the program \lq\lq Chaos, Order and Patterns; Aspects of
Nonlinearity", project No. CRG 890482.
This work is also supported by the Natural Sciences and Engineering Research
Council of Canada, les Fonds pour las Formation de Chercheurs et l'Aide a
la Recherche de la Province du Qu\'ebec, and by the Supercomputer
Computations Research Institute, which is partially funded by the U.S.
Department of Energy contract No. DE-FC05-85ER25000. All the
calculations reported here have been performed on the 64k-node
Connection Machine at SCRI. We thank Maxi San Miguel,
Mike Kosterlitz, Bertrand Morin and Emilio Hern\'andez-Garc\'\i a
for useful discussions.

\newpage

\figure{\label{fi:1}} Curvlinear coordinate system.  The points on this
figure are zeros of $\psi$ at $t=10^4$ for $\epsilon=0.25$ and $F'=0.05$.
The thick solid line was fit to the a line defined
by $\psi=0$. The
thin lines were calculated by the equation $\vec{r} =
\vec{R} + m\pi/k_{0} \hat{n}$,
were, $\vec{R}$ is the position of the thick
solid line, $\hat{n}$ is the normal to
the solid line and $m= \pm 1, \pm2, \cdots$.

\figure{\label{fi:2}} Portions of size $100 \times 100$ of typical
configurations
obtained.  The large and small circles correspond to the large and small
(see text) numerical simulations conducted at $F'=0.075, 0.05$
and $0$, and $F' = 0.025, 0.065$ and $0.1$ respectively.
The configurations labeled isotropic, nematic and smectic correspond to
$F'=0.075, 0.05$ and $0$ respectively. In all these plots
the lines drawn are the lines of $\psi(\vec{r})=0$.

\figure{\label{fi:3}} Sample configuration are displayed for $\epsilon=0.25$ at
various times and noise strengths. In all figures the points correspond to
$\psi>0$.
The configurations shown in (a) and (b) correspond to $t=2500$ and
$10^4$ respectively for $F'=0.075$.
In (c), (d) and (e), the configurations correspond to
$t=10^2, 10^3$ and $10^4$ respectively for $F'=0.05$.  A similar
time sequence is shown in Figs. (f), (g) and (h) for $F'=0$.

\figure{\label{fi:4}}  In Fig. (a), $S(k,t)$ is displayed
for $F'=0.075$ at $t=700$.  In this figure the points correspond
to the numerical result and the solid line is a fit
to $A/(B+(k^2-k_{0}^2)^2)$. In Figs. (b) and (c)  $S(k,t)$ is shown at
$F'=0.05$ and $0.0$ respectively. In both these figures, the curves
(from bottom to top at $k=1$) correspond to $t=10^2$, $10^3$ and $10^4$.

\figure{\label{fi:5}} A comparison of the one point distribution
function for $F'=0, 0.05$ and $0.075$
at $\epsilon=0.25$.  The solid, dotted and dashed lines correspond to
$F'=0$ at $t=10^4$, $F'=0.05$ at $t=10^4$ and
$F'=0.075$ at $t=700$ respectively.

\figure{\label{fi:6}} The dynamic scaling of $S(k,t)$ is
shown for $F'=0$ and
$0.05$ in Figs. (a) and (b) respectively. In
both figures the open squares, crosses
and solid circles correspond to $t=10^2, 10^4$ and $10^5$ respectively.

\figure{\label{fi:7}}
The amplitude $A(t)$, width $w(t)$ and first five moments
$m_n(t)$ of $S(k,t)$ as
defined in the text are displayed for $F'=0$ and $F'=0.05$ in
Figs. (a) and (b) respectively. The solid lines are
presented as guides to the eye and all have
slope $x=1/5$ in Fig. (a) and $x=1/4$ in Fig. (b).


\begin{references}
\bibitem[23]{re:fo2} Smaller grid spacings are needed to have the
correct spectrum of fluctuations at large $k$, but we have seen no
appreciable differences in $S(k,t)$ around $k=k_{0}$ in the nematic
region. In the disordered region or below threshold a large grid spacing
would lead to overestimating the intensity of the fluctuations, as
measured, for example, by $ \int d \vec{r} \psi^{2} (\vec{r},t)$.
\bibitem[24]{re:waffle} In order to obtain an estimate of $v$ from
reference (11) it was assumed that the high temperature values
of the constants $K_1$ and $K_3$ could be used.
\end{references}
\end{document}